\documentclass[epjST]{svjour}
\usepackage{graphics}
\usepackage{bm}
\begin{document}
\title{Liquid-crystal enabled electrophoresis:
Scenarios for driving and reconfigurable
assembling of colloids}
\author{S. Hern\`andez-Navarro\inst{1,3} \and P. Tierno \inst{2,3}
\and J. Ign\'es-Mullol\inst{1,3} \and F. Sagu\'es\inst{1,3} 
\mail{f.sagues@ub.edu}%
}                     

\institute{Department of Physical Chemistry, Universitat de Barcelona, 
08028 Barcelona, Spain \and
Department of Structure and Constituents of Matter, Universitat de Barcelona, Diagonal
647, 08028 Barcelona, Spain \and
Institute of Nanoscience and Nanotechnology, IN2UB, University of Barcelona, Spain}
\date{Received: date / Revised version: date}
%
\abstract{
We demonstrate several examples of driving and steering of
colloids when dispersed in nematic liquid crystals. The driving mechanism
is based on the principle of nonlinear electrophoresis which is
mediated by the asymmetry in the structure of the defects that the
inclusions generate in the host elastic matrix. The steering mechanism
originates in the photoactivation of the anchoring conditions of the
nematic liquid crystal on one of the enclosing plates. As experimental
realizations we first review a scenario of water microdroplets being
phoretically transported for cargo release and chemical reaction. Steering
is illustrated in terms of the reconfigurable assembly of colloidal
particles, either in the form of asters or rotating-mills, commanded by
predesigned patterns of illumination.}
\maketitle
\section{Liquid crystals as dispersing media for colloids}
One of the most distinctive properties of soft matter systems is that they are extremely
labile, and as such they may be readily controllable by means of appropriately chosen
external fields. However, this capability is useless if one does not guarantee a
sufficient control over the elicited responses, particularly when these responses are
prone to be highly degenerated. This is particularly true when referring to two of
the most studied categories of soft materials: liquid crystals and colloids. In particular,
interest has increased enormously in recent years to bridge the potentialities of
both systems through the use of liquid crystals as dispersing media for solid or liquid
colloidal inclusions. Such composite systems, where we benefit of the discrete nature
of the dispersed components and, at the same time, of the orientational properties
of the liquid crystal medium, constitute nowadays familiar mixtures under intense
scrutiny to unveil new fundamental concepts and original applications~\cite{1,2,3}. In particular,
the capacity of the elastic and anisotropic host matrix to mediate interactions 
between suspended inclusions was already discovered many years ago~\cite{4}, and has been
profusely used since then~\cite{5,6,7,8,9,10,11,12,13,14}.

In contrast to research on liquid crystal dispersions of sub-micron or nano-scale
particles~\cite{15,16}, we concentrate in what follows on systems that contain colloidal
inclusions at the micrometer scale, which enable a real discrete control and observation
\begin{figure*}[t]
\resizebox{\columnwidth}{!}{\includegraphics{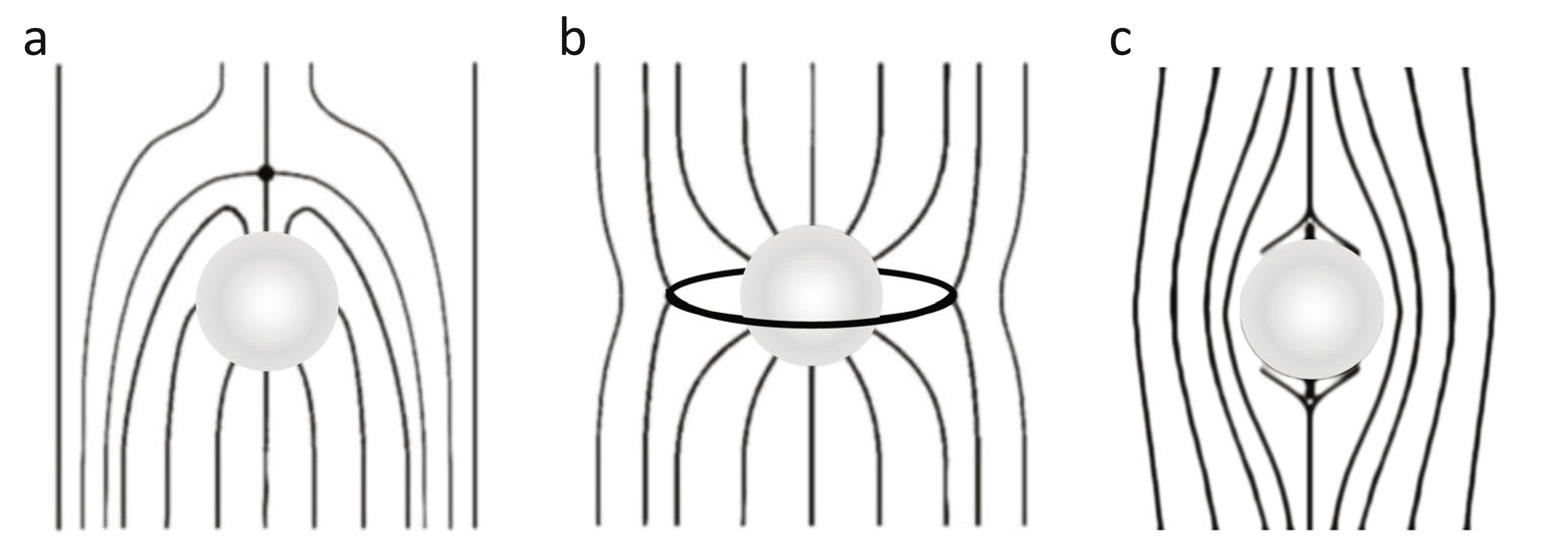}}
\caption{Different defect structures for inclusions dispersed in nematic liquid crystals.
a) Hedgehog configuration; b) Saturn-ring disclination; c) Double-boojum structure. The
first arrangement has a dipolar symmetry, whereas the other two feature distortions of
quadrupolar nature.}
\label{fig:1}       
\end{figure*}
of the dispersed components. It is also worth emphasizing at this point that, in spite
of very recently investigated dispersions prepared with lyotropic (i.e. water-based)
liquid crystals~\cite{17,18}, all what we are going to review in this contribution refers to
thermotropic oily materials in their nematic phases~\cite{19}.

Nematic liquid crystals (NLCs) are complex fluids mostly used in display technologies.
They are characterized by rod shaped organic molecules which tend to
align their long axis on average along a common direction, called the director field.
The behavior of colloids dispersed in NLCs is largely determined by considering
the state of the elastic LC matrix around the dispersed inclusion. In this respect,
the most important characteristics is the anchoring conditions of the liquid crystal
at the surface of the colloid. The simplest case is that of a spherical solid particle,
around which the nematic material may adopt either a tangential (parallel to
the surface) or a homeotropic (perpendicular to the surface) arrangement, depending
on the particular system and surface preparation. In either case, placing a colloidal
unit into an otherwise aligned nematic phase leads to topological frustration
and to the appearance of defects (i.e. singularities in the distribution of the director
field), that are tightly bound to the inclusion. For homeotropic anchoring, two
main types of defects may arise: a point-defect known as a hyperbolic hedgehog
or a disclination loop extending as an equatorial band and known as Saturn-ring~\cite{20,21,22}. For tangential anchoring, the defect structure is that of a double-boojum~\cite{20}. Schematic representations of these defect configurations are presented in Fig. 1.
\begin{figure*}[h]
\resizebox{\columnwidth}{!}{\includegraphics{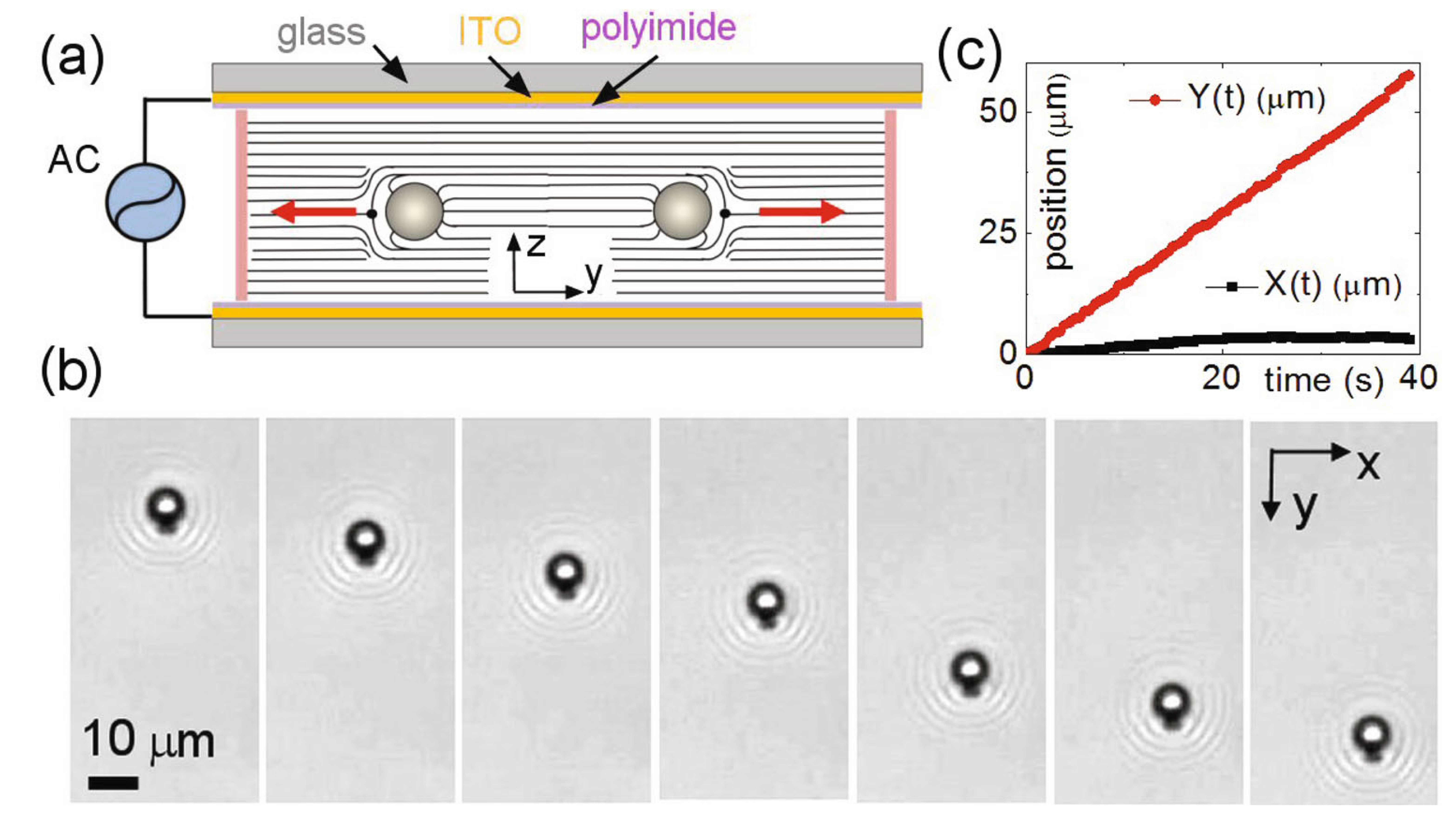}}
\caption{(a) Schematics of the experimental cell with water microdroplets dispersed in a
nematic liquid crystal contained between two transparent glass electrodes functionalized to favor planar anchoring. The water droplets are propelled by the mechanism of LCEEP
(see text). (b) Optical micrographs taken every $4.5$s of a water droplet having a $6.5 {\rm \mu m}$
diameter and moved at a speed of $1.7 {\rm \mu m s^{-1}}$
by an AC field of amplitude $E = 0.7 {\rm V \mu m^{-1}}$
and frequency, $f =10$Hz. (c) Positions ($x,y$) versus time of the droplet shown in panel (b).
Image reproduced with permission from ~\cite{44}. Copyright Royal Society of Chemistry 2013.}
\label{fig:1}       
\end{figure*}
Of singular importance on what follows is the elucidation of the symmetries of
these defects. In this respect, Poulin et al., already pointed out in their seminal
paper ~\cite{4} the convenience to map the colloidal-induced elastic distortion of the far
field distribution of the nematic director onto an electrostatic context. According
to this analogy, a (ideal) spherical colloid with its corresponding hedgehog defect
acts as an electrostatic dipole, whereas the structures of Saturn-rings and doubleboojums
are better understood as giving rise to quadrupolar distortions of the director
field.

\section{From non-linear electrophoresis to liquid-crystal enabled
electrophoresis}

Another more recently investigated feature of nematic colloids, this term used as a
short denomination for colloidal dispersions in a continuous nematic phase, is the
possibility to observe non-trivial manifestations of phoretic driving of the dispersed
inclusions under the application of electric fields. Indeed, what makes this possibility
particularly striking is that, contrarily to normal phoresis under steady fields, the
motion is here induced by oscillating electric fields. This confers to the phenomenon a
distinctive non-linear character that justifies the use of the abbreviated denomination
non-linear electrophoresis~\cite{23}. An alternative way to qualify this electrically-driven
transport, more appropriate to its physical origin, is known as induced-charge electroosmosis
~\cite{24}. Focusing on liquid crystal based systems, we will employ on what
follows the composed form liquid-crystal enabled electrophoresis LCEEP, as was recently
coined by Lavrentovich et al.~\cite{25}.

For normal electrically induced phoresis, particles are transported under uniform
DC fields with a velocity that depends linearly on the applied field. Obviously this
means that oscillating fields which average to zero over a period would cause a zero
displacement. However the possibility to use alternate rather than steady driving
modes has the advantage that concurrent electrochemical processes, that take place
in the normally aqueous medium employed in electrophoretic cells, are avoided. In
fact, nonlinear versions of electrophoresis were first reported for isotropic fluids more
than twenty years ago~\cite{26}, as mentioned in the recent review~\cite{25}, and were exhaustively
analyzed in some theoretical papers by Bazant and Squires~\cite{24}. In this latter
study, the main conclusion is that a nonlinear electro-osmotic slip occurs when an applied
field acts on the ionic charge it induces around a polarizable surface. Notice from
this very first statement that nonlinear electrokinetic phenomena apply to charged
and non-charged inclusions as well, in striking contrast with normal electrophoresis.
As a matter of fact, the possibility to further break the symmetries of the resulting
flow and obtain direct motion for Janus-like particles (metallo-dielectric) was first
published by Velev et al.~\cite{27}.

In the context of liquid crystals, the breaking of the fore-aft symmetry that guarantees
direct transport may be achieved even for perfectly spherical particles, bearing
in mind our above considerations relative to the symmetries of the nematics director
field around the inclusions. This is quite evident for defects with dipolar symmetry,
but also applies to distorted quadrupolar symmetries. The latter situation would occur,
for instance, in the case of a parallel anchoring of the liquid crystal material
wrapping around non-perfectly symmetric, i.e., anisometric, particles. Both possibilities
have been largely exploited in our recent experiments that will be reviewed in the
second part of this contribution. Notice finally that a quadratic dependence of the
phoretic velocity with respect to the applied field has another striking implication:
the dependence of the velocity on the applied electric field, both vectorial quantities,
must be indeed of a tensorial nature. This, in turn, permits that the direction of motion
and that of the electric field are non-necessarily parallel~\cite{23}. A general overview
on the different possibilities involved in the transport of particles in liquid crystals is
provided in ~\cite{25}.

\section{Phoresis of nematic colloids: An active matter-based perspective
and applications in materials science}

Active matter is a presently very celebrated keyword under which we recognize many
forms of elementary living matter, altogether with remarkable non-living soft matter analogs~\cite{28}. A defining feature commonly exhibited by the units composing these
discrete systems is their ability for autonomous or driven motion. More precisely, we
should properly reserve the term active for the first kind of (self-driven) behavior and
employ the term actuated to refer to the second situation, which is in fact what applies
to our electrophoretic colloids dispersed in a liquid crystal material. Under whatever
acception, much attention has been devoted to elucidate the mechanisms behind
complex multiparticle phenomena that self-organize into characteristic patterns with
long range order, either of static or dynamic nature. In this respect, terms like bands,
asters, vortices are very much familiar to practitioners in this field. Diverse aspects of
this topic are examined in this Proceedings volume while a comprehensive update is
provided in the review by Marchetti et al.~\cite{28}. Although experimental realizations are
becoming more and more impressive particularly when restricting to a bio-inspired
context, see ~\cite{29,30,31} for striking observations of active self-organization in a biophysical
context, we feel that theoretical and numerical work is leading the activity in the field.
We thus claim for the need to provide the interested community with experimentally
robust realizations of new scenarios of self-organized patterns in soft-matter systems,
either active or driven. More important, as it is going to be much emphasized on
the last part of this contribution, is the further need to gain control on the assembly
process and to demonstrate the eventual possibility to reconfigure these aggregates
at will.

We would like to mention another completely different and more materials based
perspective from where our experimental research on phoretic colloids could
be addressed. The assembly and transport of colloidal entities, such as particles,
droplets or microorganisms have direct applications in fields such as photonics ~\cite{32},
lab-on-a-chip technologies ~\cite{33} and biomedicine ~\cite{34}. During this last decade the focus
on self-assembling started to shift from equilibrium characteristics to dynamic
aspects~\cite{35}, seeking enhanced functionalities of the resulting materials~\cite{36}. Optical
trapping techniques are often employed to achieve direct control over the placement
of colloidal inclusions, and holographic tweezers allow to extend such control to a few
hundreds of particles~\cite{37}, although this technique is limited by the field of view of
the optical system. Going beyond earlier work on physically and chemically activated
colloids~\cite{38,39,40}, a recent realization explores the idea of phoretic colloids driven by
osmotic pressure imbalances~\cite{41}. These results have opened new possibilities for reconfigurable
self-assembly, enabling the massive transport of inclusions~\cite{42}, although
the precise and selective control on such dynamical patterns lacked experimental
evidence.

\begin{figure*}[t]
\resizebox{\columnwidth}{!}{\includegraphics{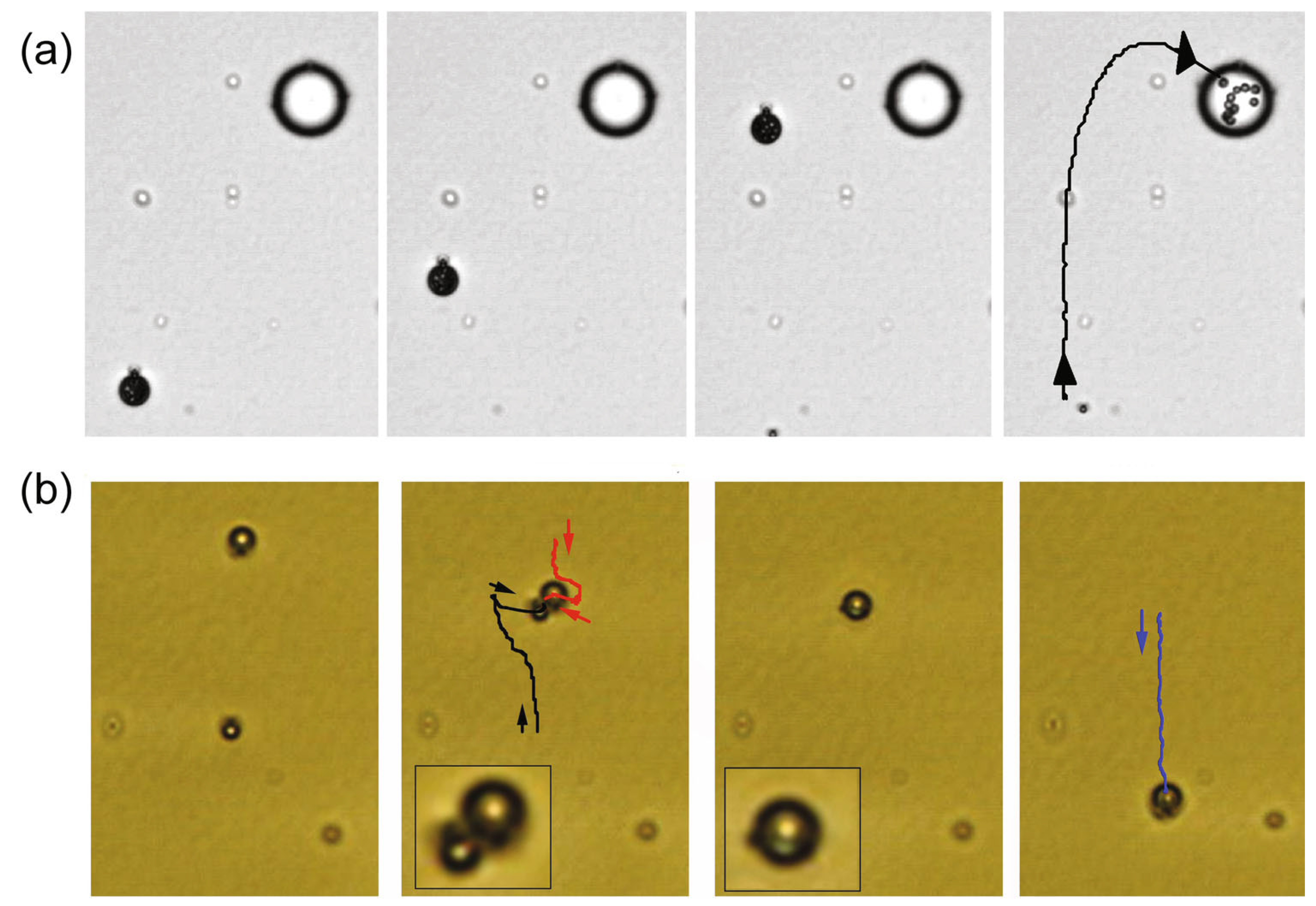}}
\caption{(a) Sequence of images showing the coalescence of a microdroplet of $7.3 \rm{\mu m}$ diameter
filled with $11$ smaller polystyrene particles with a large droplet of diameter $18 \rm{\mu m}$ bearing
a Saturn-ring defect ($E = 0.70 \rm{V \mu m^{-1}}$
, $f =10$Hz). Time interval between images is $17.2$s.
(b) Microscope images showing two water microdroplets with diameters $2.7$ and $3.7\rm{\mu m}$
driven in opposite directions by an AC field of the same characteristics. Droplets contain
separated reactants, respectively potassium ferrocyanide ($0.2$M) and ferric ions ($0.3$M). The
microdroplets approach and coalesce into a larger one forming Prussian blue as a precipitate.
Image reproduced with permission from in~\cite{44}. Copyright Royal Society of Chemistry
2013.}
\label{fig:1}       
\end{figure*}

\section{AC electrophoresis of microdroplets in liquid crystals: Transport
and reaction}

The second part of this contribution is prepared to briefly summarize our recent research
on the phoresis of inclusions dispersed in nematic liquid crystals. It is itself
divided into two sections. The first one is devoted to the transport of (water) microdroplets,
while we reserve the second part to present a new development that permits
the reconfigurable assembly of colloidal particles into clusters (swarms).

Water-based microemulsions prepared in oil phases are of fundamental importance
in chemical and analytical sciences, due to the possibility of encapsulating and
delivering chemical compounds otherwise immiscible in the dispersion medium. The
possibility to remotely transport microdroplets is thus singularly appealing. Electrically
induced phoresis is undoubtedly the first considered alternative, and particularly
its non-linear version based on the use of oscillating electric fields offer great advantages
as mentioned previously in Sect. 2

\begin{figure*}[t]
\resizebox{\columnwidth}{!}{\includegraphics{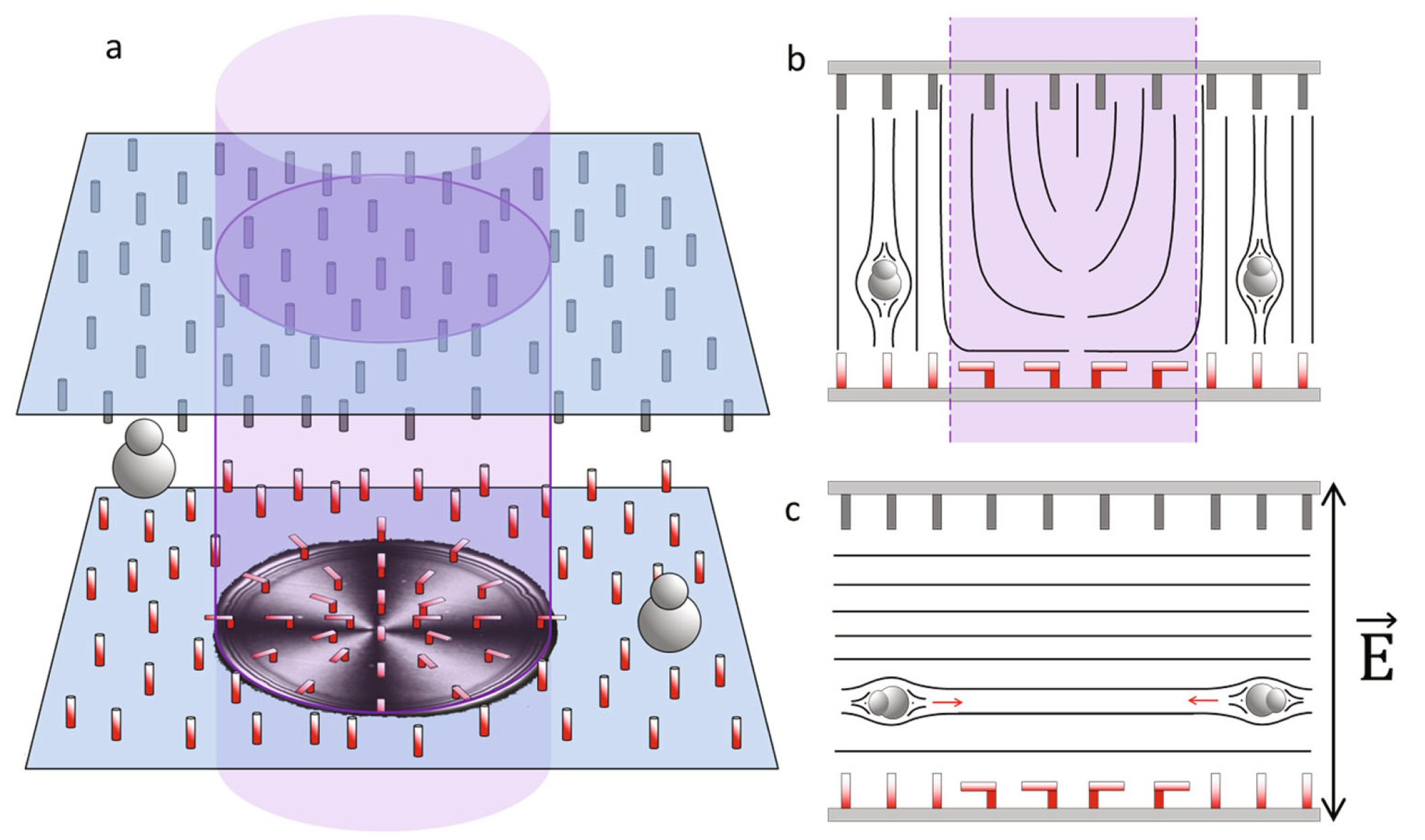}}
\caption{(a) Scheme of the experimental cell as a radial pattern is imprinted using UV light
($365$nm). The bottom plate is functionalized with an azosilane derivative, and the upper
plate presents homeotropic anchoring. (b) Scheme of a transversal cut of the experimental
cell. (c) Scheme of a transversal cut of the experimental cell, upon application of a sinusoidal
electric field. Red arrows indicate the dominant direction of motion of the anisometric
particles.}
\label{fig:1}       
\end{figure*}

Our experimental system, thoroughly described in Hern\'andez-Navarro et al. ~\cite{44},
permits to transport water microdroplets dispersed in a nematic liquid crystal with
negative dielectric anisotropy, i.e. it allows to observe phoretic motion perpendicular
to the applied electric field. Additionally, we decouple in this way AC electrophoresis
from any residual linear DC contribution. As reported in the reference mentioned,
we were able to demonstrate not only droplet motion but to show in addition that
these droplets can be used as microreactors to transport sub-micrometric particles or
to mix tiny volumes of chemicals.

The used cell was composed of two 0.7$mm$ thick microscope slides of size
$15 \times 25 \rm{mm^2}$ coated with a layer of indium-tin oxide (ITO). The two slides were
cleaned, dried and further chemically treated to obtain a planar alignment of the
liquid crystal on the bounding plates. These were separated by a spacer of nominal
thickness $23$ microns and glued together with the ITO layers facing inwards. Dispersions
of aqueous microdroplets (from $1$ to $20$ micron diameter sizes) were prepared
by vortex agitation using MLC-7029 as a nematic phase, using as stabilizing agent
sodium dodecyl sulphate (SDS). This protocol guarantees a homeotropic alignment
of the nematics on the droplets surface. The experimental cells were filled by capillarity.
Sinusoidal electric fields were applied by using a function generator, within a
range of $0$ to $30$ volts peak-to-peak, while the applied range of frequencies varied from
$0$ to $100 $Hz. Experimental observations were performed with an optical microscope,
and images were captured with a recording camera controlled with the appropriate
software AVT SmartView 1.10.2, and further treated using software packages ImageJ
and IgorPro. For more details see the original paper ~\cite{44}.

\begin{figure*}
\resizebox{\columnwidth}{!}{\includegraphics{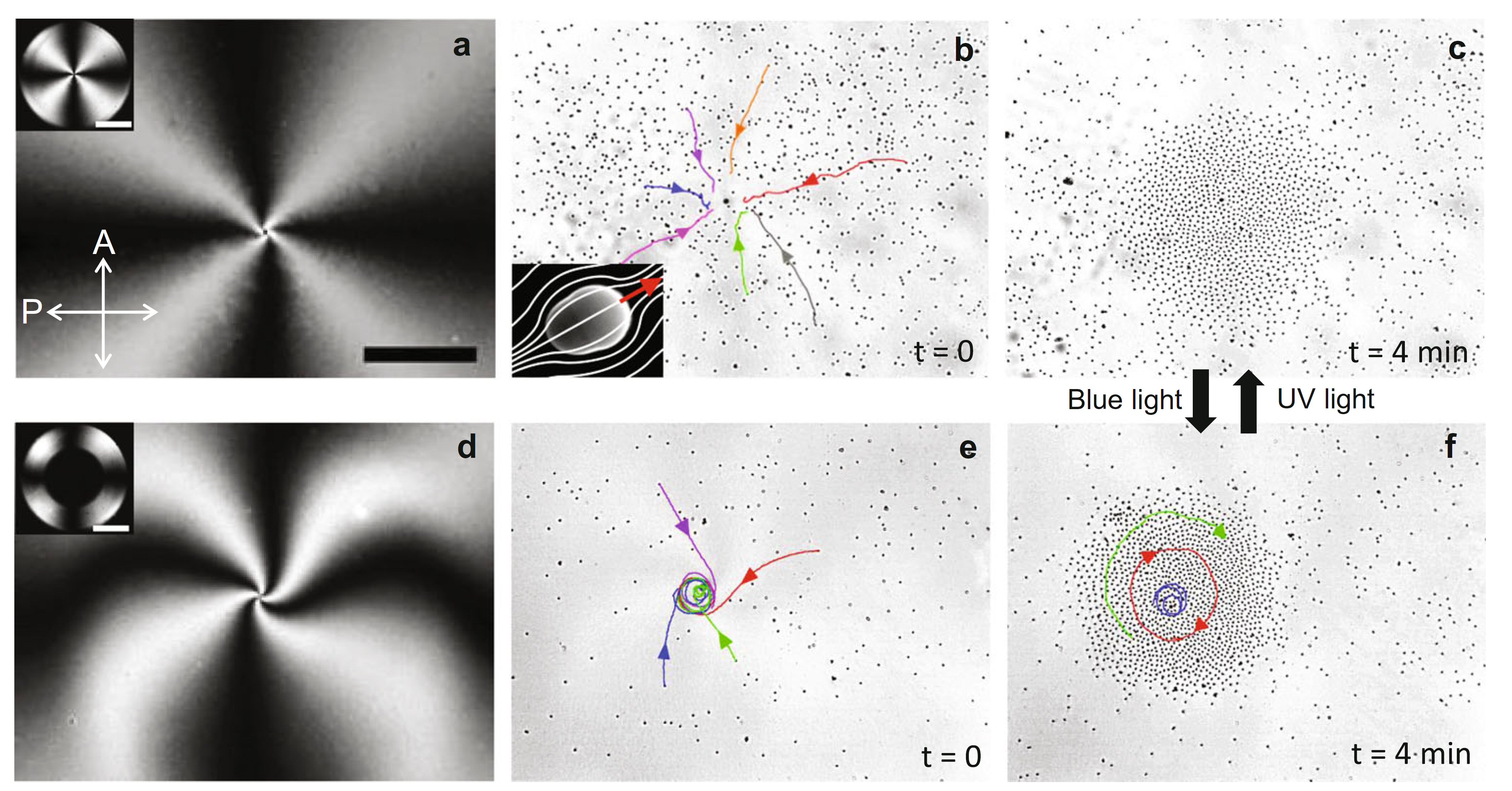}}
\caption{Top (a-c) row of images illustrates the formation of a colloidal aster. Bottom (d-f) row
corresponds to the assembly of a rotating mill-like cluster. (a) and (d) are images between
cross polarizers of the imprinted texture leading to a cross (a) or a spiral (d) attracting pole.
The inset shows the planar photoaligned circle (inset (a)) or corona (inset (d)) prior to the
application of the electric field. The applied electric field in both cases has an amplitude
of $0.87 V\rm{\mu m^{-1}}$ and a frequency of $10$Hz. Trajectories followed by the several particles are
superimposed to the images. Inset in (b) shows a S.E.M. image of a single pear-shaped
particle ($3 \rm{\mu m}\times 4 \rm{\mu m}$). The two cluster modes can be interconverted by suitable irradiation
protocols as indicated. The scale bars are $200 \rm{\mu m}$ for all  images and $500 \rm{\mu m}$ for the insets. Image adapted with permission from~\cite{46}. Copyright Wiley 2014.}
\label{fig:1}       
\end{figure*}
\begin{figure*}
\resizebox{\columnwidth}{!}{\includegraphics{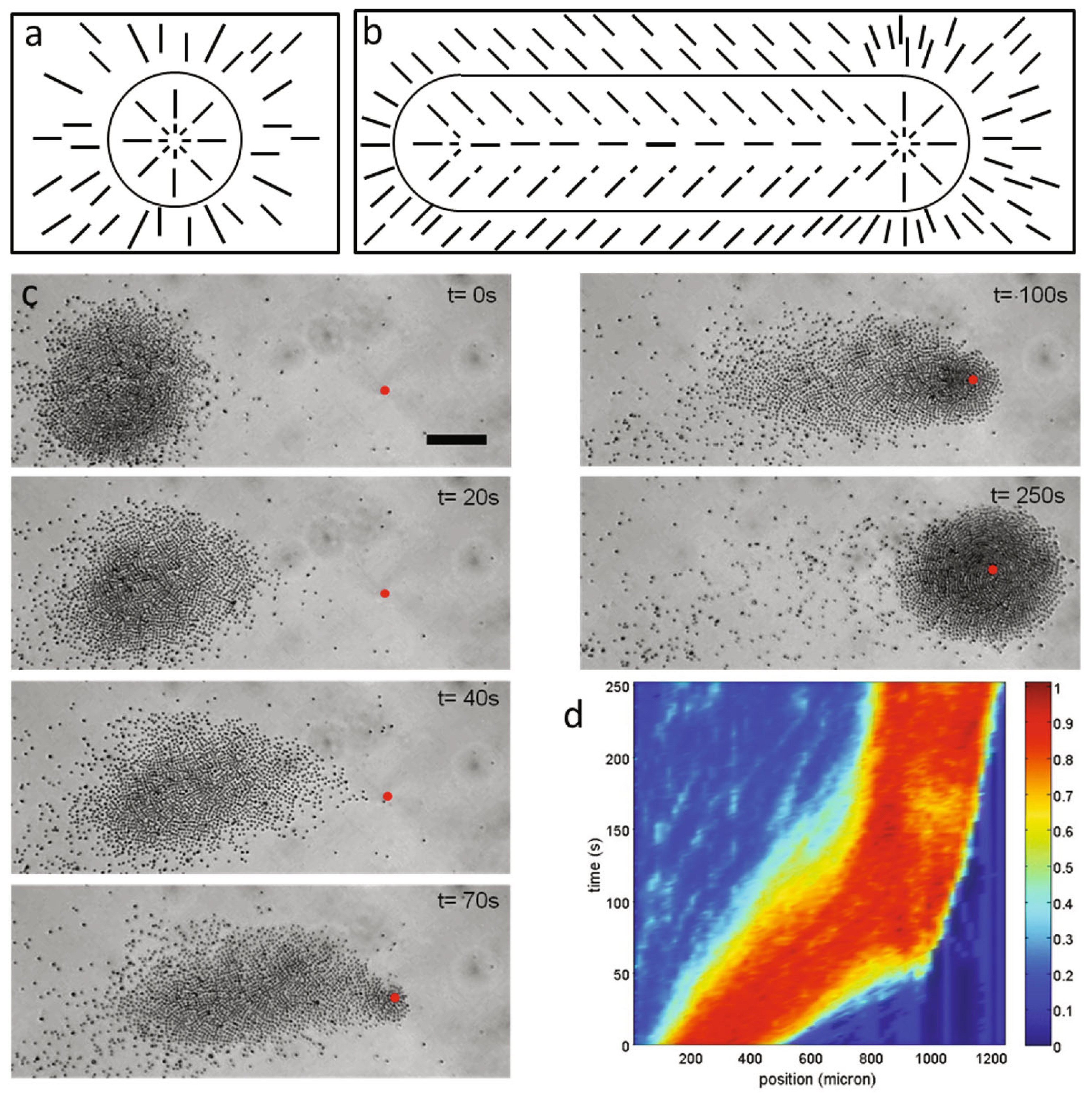}}
\caption{Image sequence of a particle swarm traveling across the LC cell due to in situ reconfiguration
of the NLC field. (a) and (b) are schematic representations of the LC orientational
field right at starting and during relocation of the cluster. (c) The photoaligned spot initially
centered with the cluster is moved $600 \rm{\mu m}$ to the right, as indicated by the red spot. The
applied sinusoidal electric field has an amplitude of $0.74V {\rm \mu m^{−1}}$ and a frequency of $10$Hz.
(d) The chart is a space-time plot visualizing the swarm dynamics during the experiment.
The normalized average particle density across different positions is color encoded, with dark
blue corresponding to particle depleted and dark red to particle rich regions. Scale bar marks
$150 \rm{\mu m}$.}
\label{fig:1}       
\end{figure*}

Figure 2(a) shows the experimental cell. In the scenarios we are reviewing here
we found that the defects were predominantly of the hedgehog type, while only a
small fraction of inclusions (less than 1\%) featured Saturn-ring defects. For isolated
droplets it is known~\cite{45} that the hedgehog is the usual defect configuration for large
inclusions, whereas Saturn-ring distortions are preferred in smaller ones. However the
situation is more complicated for confined systems, and in this case Saturn-ring defects
are, in general, limited to large droplets, as observed in our experiments. Motion
was always observed in the direction towards the point defect as clearly observed in
Fig. 2(b). Motion of the droplet in relation to the location of the point defect may,
however, depend on the used liquid crystal, as reported in~\cite{44}. The measured velocities
are typically on the range of a few microns per second, featuring the expected
quadratic dependence on the amplitude of the electric field and a peak at intermediate
frequencies (tens of Hertz).

To further demonstrate the possibilities of this phoretic motion, a localized cargo
release operation is shown in Fig. 3(a). A microdroplet loaded with polystyrene particles
is driven towards a larger droplet bearing a Saturn-ring defect, and thus stays
at rest under the applied electric field. The larger droplet elastically attracts the
small one, and the colloidal cargo is released by droplet coalescence. In Fig. 3(b),
two droplets loaded with specific reactants and featuring antiparallel configurations
of their respective dipolar defect configuration attract each other, and their coalescence
triggers a chemical precipitation reaction (see original paper~\cite{44} for further
details).

\section{Reconfigurable swarming}

Reconfigurability is a genuine property of living active matter. As laboratory-model
systems, collections of driven colloids have a potential for large-scale addressability.
Here we review our very recent research that indeed demonstrate real-time reconfigurable
clustering of phoretic colloids dispersed in thin nematic layers in response to
illumination patterns. Swarms of particles are shown to reversible organize in submillimeter
ensembles that can be individually or collectively addressed to change their
position or dynamic behavior. The strategy we follow permits to separate particle
steering, achieved through photoelastic modulation of the host nematic, and particle
driving realized through LCEEP. A complete account of these experiments can be
found in Hern\`andez-Navarro et al.~\cite{46}.

We used the same nematic liquid crystal as in the previously reported experiments,
thus favoring motion of the dispersed colloids perpendicular to the applied electric
field. However, the chemical treatments of the confining plates was different in this
case and, in fact they were specifically aimed at achieving the purposed control on the
assembly of the dispersed particles. One of the plates was chemically functionalized
with an azosilane photosensitive self-assembled monolayer that allows to alternate between
perpendicular (homeotropic) and tangential (planar) anchoring of the nematic~\cite{47}. The counter plate was coated with a polyimide compound to guarantee strong
and permanent perpendicular contact. Without any external influence, such boundary
conditions lead to a uniform homeotropic texture within the whole cell. By irradiating
with UV ($365$nm) light from an incoherent source the azosylane is forced to adopt
the cis configuration, and consequent planar boundary conditions are attained. The
azosylane is easily reverted to trans form using blue ($455$nm) illumination, and corresponding
homeotropic anchoring is regained. Light-induced reorientations typically
take a few seconds, a much shorter time scale than that set by particle motion.

As colloidal inclusions we used pear-shaped microparticles made of polystyrene,
a material that promotes planar orientation of the NLC on the particle surface. The
chosen particle shape guaranteed a distorted quadrupolar symmetry arising from the
defect distribution around the particles. In the absence of irradiation or electric field,
particles align perpendicular to the cell plates, following the uniform director field.
Illuminating the cell for a few seconds with a spot of UV light forces the NLC in contact
with the azosylane-treated surface to transit to a planar configuration by locally
adopting a local splayed (radially-spread) texture emanating from a central defect.
Application of an external AC field makes the bulk NLC to adopt the splay configuration
that now extends for several millimeters, well beyond the area of the irradiated
spot, thanks to the homeotropic anchoring degeneracy relative to any planar direction.
This configuration is stable for days under AC field, well past the half-life for
thermal relaxation of the azosylane film, which is about $30$ minutes. The region with
radial alignment will be the basin of attraction for dispersed particles, which tumble
instantaneously following the NLC director so that their long axis lays, on average,
parallel to the cell plates. Simultaneously the LCEEP sets the particles into motion at
a constant speed. All particles moved following the local NLC director, with roughly
half of them being attracted by the photoinduced radial defect and the rest being
repelled from it. This is direct consequence of the random tumbling of the particle
following the reorientation of the liquid crystal matrix. Most of the particles propel
with the large lobe ahead, as reported in~\cite{46}. See Fig.4 for a combined schematics
of the different aspects of the optical and electrical forcing of the cell.

During the experiment, particles accumulate as they follow the nematic field lines.
A growing aster-like cluster was typically observed after the progressive arrival and
subsequent jamming of the particles leading to high density assemblies, see Fig. 5
(panels a, b, c). We could easily switch to a dynamic structure, a rotating mill-like
cluster, by profiting from the elastic properties of the nematic material (smaller value
of the bend elastic constant with respect to the splay counterpart) and the fact that
the particles are totally slave to the director. The experimental protocol proceeds by
erasing the central region of an imprinted UV area with a smaller spot of blue light,
prior to the application of the electric field. Following the inverse isomerization to
the trans form of the azosylane anchoring derivative, the NLC director features a
homeotropic configuration both outside and inside a corona with planar alignment.
Upon application of the AC field, the NLC director adopts a degenerate planar alignment
both inside and outside the ring. The energy cost of the large splay distortion
inside the ring prompts the director to adopt a spiral bend-splay texture and director
field lines conform to a spiral-like geometry. As a consequence particles assemble into
a rotating mill structure, preceding around the central defect with a constant linear
velocity, see Fig. 5 (panels d, e, f). Both assembly modes can be reversibly interconverted
in real time via the photoativation control just described (see original reference~\cite{46}). It is worth empasizing that, as expected, there is no preferent sign in the cluster
rotation, since there is no chiral influence in the system. In this respect particle aggregation
here is very different with respect to assembling of particles under a forced
imposed vortical flow (see for instance~\cite{48}).

The reversibility and quick response of the photoalignment layer enables straightforward
cluster addressability. A preformed aggregate of arbitrary size, either aster
or vortex-like, can be relocated to a pre-designed place anywhere within the experimental
cell with minimum dismantlement of the cluster structure by changing the
location of the UV spot. An example of this process is shown in Fig. 6. After blocking
the LCEEP mechanism by increasing the field frequency above 50 Hz, the center of
attraction is translated 600 μm. Once LCEEP is reactivated, the swarm of particles
moves towards the new position developing a leading edge around which the particles
assemble. The space-time plot illustrates the resulting collective behavior. Alternatively,
by the same principle one can imprint predesigned arbitrary paths connecting
distant locations inside the cell or draw circuits with complex topologies as a simple
way to accumulate colloidal swarms in the irradiated area and further entrain them
collectively.

We thank Patrick Oswald for the polyimide compound. We acknowledge financial support
by MICINN (FIS2010-21924C02, FIS2011-15948-E) and DURSI (2009 SGR 1055). S.H.-N.
acknowledges support through an FPU Fellowship (AP2009-0974). P.T. further acknowledges
support from the ERC through the starting grant DynaMO (335040) and from the Ramon
y Cajal program (RYC-2011-07605).

\end{document}